# Metrics for Evolution of Aspect Oriented Software

S. Senthil Velan
Department of Computer Science and Engineering
SSN College of Engineering
Old Mahaballipuram Road
Kalavakkam Tamil Nadu INDIA
senthilvelan@ssn.edu.in

Dr. Chitra Babu
Department of Computer Science and Engineering
SSN College of Engineering
Old Mahaballipuram Road
Kalavakkam Tamil Nadu INDIA
chitra@ssn.edu.in

*Abstract*— **Aspect Oriented Software Development (*AOSD*) is a promising methodology which provides powerful techniques to improve the modularity of the software by separating the cross-cutting concerns from the core functionality. Since evolution is a major requirement for the sustainability of any software, it is necessary to quantitatively measure its impact. In order to quantify, it is essential to define metrics that will capture the evolution of Aspect Oriented (AO) software. It is also necessary to compare the metric values of various versions of software to draw inferences on the evolution dynamics of AO software. This needs identification of artifacts that were added, deleted or modified across versions and study the consequence of these types of changes. This paper defines a new set of metrics for measuring the evolution of Aspect Oriented software. As a case study, an aspect refactored software, *AJHotDraw* has been chosen and its four versions have been analyzed for their capability to evolve over time.**

*Keywords-AOSD, Software Evolution and Software Metrics*

## I. INTRODUCTION

Complex large scale software often consists of code that realizes several different concerns (features). Such software contains code that are related to concerns such as transaction management, exception handling and other non-functional requirements apart from the code to realize the primary concern that deals with the core business logic. These concerns interact with the core concern and are called as cross-cutting concerns. For any software, modularity is very important to understand the flow of execution. Although, object oriented programming introduced the concept of classes which encapsulate the data and its associated methods, ensuring that the data members can be accessed only by the member functions, it suffers from limitations such as code tangling and code scattering.

Aspect Oriented Programming (AOP) [1] is a paradigm that overcomes these limitations by modularizing the cross-cutting concerns through specification of a new unit of abstraction called as *Aspect*. According to Kiczales [1], an aspect is defined as "a modular unit of cross-cutting implementation. It is defined very much like a class, and can have methods, fields, constructors, initializers, named pointcuts and advice". AOSD enhances the separation of concerns, improves the software reusability and reduces complexity. It also provides greater flexibility during the development of software and eases the evolution of software.

Software evolution refers to the study and management of the process of making changes to software over a period of time. These changes can be of three types namely, adding a new functionality, deleting an old functionality, and modifying an existing functionality in the software. It can be stated that evolution is an essential feature for any sustainable software.

There are several stages in the development of software. Initially, the knowledge about the software needs to be gathered, such as the domain of the application and the user requirements. The next stage deals with the selection of the appropriate data structures, algorithms, architecture, and operating environment. This stage is very vital for the subsequent phases of evolution. The first version of the software is now developed using the artifacts obtained from the former stages. The requirements of the user and the operating environment are not static. Hence, the development of the software does not end with its first release. Modifications are done to the software ensuring the preservation of its architectural integrity. The following are the typical set of scenarios where the software evolution usually occurs:

- Since all the requirements cannot be clearly specified in the initial stages of software development, there is a need for the change in requirements to be reflected in the subsequent versions of the software.

- Usually in the first version of the software, the basic sets of functionalities are implemented. The extended capabilities are incrementally included in the subsequent versions.

- When the bugs in the software need to be fixed, the software also evolves.

- In order to enhance the performance of the system, related features are added/deleted/modified to the existing system.

- The business environment is never static and is subject to changes for keeping itself in tune with the growth.

This paper has proposed a set of metrics which influences the software quality attribute – *Evolution*. An Aspect Oriented Software Evolution Measurement Tool (*AOSEM Tool*) has been developed for obtaining the values of the proposed set of metrics. The rest of the paper is organized as follows. Section II discusses the related work. The scope of the problem is briefly stated in Section III. Section IV proposes the new set of metrics for measuring the evolution of AO software. Section V discusses the case study, *AJHotDraw* and how the proposed set of metrics has been evaluated for its four different versions using the *AOSEM Tool*. Section VI explains the results obtained for the proposed metrics across different versions of *AJHotDraw*. This section also explains the evolution dynamics of *AJHotDraw* through the metric values. Section VII concludes and provides future directions.

## II. RELATED WORK

Turski [2] proposed that system growth can be measured in terms of number of source modules and number of modules changed. It is also observed that the system growth is generally sub-linear and slows down as the system becomes larger and complexity increases.

Kemerer [3] performed a study of software evolution which concentrated on the types of changes, costs and effort to evolve the software. He has analyzed these factors from the perspective of Lehman's laws by comparing the time series and sequence analysis of data.

In a study by Anderson and Felici [4], an avionics safety-critical system was examined. There was a great variation in the number of new services between releases. The Requirements Maturity Index *(RMI)* was used to measure the relative change in the number of requirements and is given by the formula depicted through equation 1.

$$RMI = (R_T - R_C)/R_T \qquad (1)$$

where,

$R_T$ is the number of requirements in the current release and,

$R_c$ is the number of requirements that were added or deleted or modified from the previous release.

Sai Zhang [5] used change impact analysis approach for AspectJ programs which captures the semantic differences between AspectJ program versions.

Zhao [6] has made an attempt to extend the program slicing technique that is usually used to study the change impact analysis of procedural and object oriented software onto aspect oriented software. A Dependence graph was constructed for a small code segment and the base code is program sliced to find the impact of an aspect on a given join point.

## III. SCOPE OF THE PROBLEM

A set of metrics is defined which influences the software quality attribute - *Evolution*. An Aspect Oriented Evolution Metrics Tool, *AOSEM Tool* was developed to determine the values of the defined set of metrics for the given AO software. Four versions of *AJHotDraw* were considered for the study of AO evolution. Each version is evaluated using the *AOSEM Tool* and the metric values are obtained. Finally, inferences have been drawn regarding the evolution dynamics of the AO refactored software *AJHotDraw*.

## IV. PROPOSED METRICS

A change to the software developed using aspect oriented methodology involves identifying the set of changes in aspectual elements and also the changes made in the base elements (class elements). While defining the set of metrics, deletion of entities has not been accounted since, in a typical software, deletion will usually be related to addition of new elements. Any added element would exhibit functionality that may be equivalent or advanced while comparing to its deleted counterpart. Further, since the first element of each proposed metric is the number of the entities in the current version, the number of deleted entities is already accounted for. In the four versions of *AJHotDraw,* the candidate functionalities to be encapsulated as aspects are refactored from *JHotDraw6.0*. Hence, new functionalities are not added in any of the four versions. While computing the values of metrics, if the total number of current elements is zero then the metric value is also zero. The proposed set of metrics will capture the respective changes in each entity of the software.

### A. Metrics for change in aspect elements

Aspects have a set of unique elements such as, aspect, pointcut, joinpoint, advice and introduction. Hence, the various changes that are possible and captured in an aspect code are change in aspects, change in pointcuts and change in advices.

*1) Change in Aspects(CIA)*: An aspect is similar to a class in Java and handles the encapsulation of join points, pointcuts, advices and inter-type declarations related to a particular cross-cutting concern. As the software evolves, new aspects may be added to meet the changing requirements and some may even be modified to accommodate the additions and deletions. By applying the same method used to calculate the *RMI* (equation 1), the *Aspect Maturity Index (AMI)* of each version can be obtained using equation 2. The value of *AMI* will lie between the range of 0 to 1. Using this maturity index, the change beween versions can also be caputred using equation 3 specified below:

$$AMI = (A_c - (A_a + A_m))/A_c \qquad (2)$$
$$CIA = 1 - AMI \qquad (3)$$

where,

$A_c$ is the number of aspects in the current release,

$A_a$ is the number of aspects that were added to the current release and,

$A_m$ is the number of aspects that were modified from the previous release to obtain the current release.

*2) Change in Pointcuts(CIP)*: A pointcut is designed to identify and select join points within an AspectJ program. A join point is a well-defined location within the primary code where a concern will crosscut the application such as method calls, constructor invocations, exception handlers, or other points in the program. New pointcuts may be added or the existing pointcuts may be deleted or the body of the pointcut may be modified during software evolution. Hence, by using the same method of calculating the *CIA*, the *Pointcut Maturity*

Index (PMI) can be determined using equation 4. The change in pointcuts between versions can be calculated by applying equation 5 given below:

$$PMI = (P_c - (P_a + P_m))/P_c \quad (4)$$
$$CIP = 1 - PMI \quad (5)$$

where,

$P_c$ is the number of pointcuts in the current release,

$P_a$ is the number of pointcuts that were added to the current release and,

$P_m$ is the number of pointcuts that were modified from the previous release to obtain the current release.

*3) Change in Advices(CIAD)*: An advice is a function, method or procedure that is executed when a given join point of a program is reached. It can execute at three different places when a join point is matched: before, around, and after. In each case, a pointcut must be triggered before any of the advice code is executed. Similar to the pointcut, a new advice can be added or the existing advice can be deleted or the body of the advice can be changed as the software evolves. The values of *Advice Maturity Index(ADMI)* and *CIAD* are computed using equations 6 and 7 specified below:

$$ADMI = (AD_c - (AD_a + AD_m))/AD_c \quad (6)$$
$$CIAD = 1 - ADMI \quad (7)$$

where,

$AD_c$ is the number of advices in the current release,

$AD_a$ is the number of advices that were added to the current release and,

$AD_m$ is the number of advices that were modified from the previous release to obtain the current release.

### B. Metrics for change in base elements

A class in the base code encapsulates two types of elements – data members and the corresponding methods.

*1) Change in Classes(CIC)*: A class can be added, deleted or modified during software evolution. Initialization of a class and invocation of methods are qualified joinpoints while modeling an application using aspects. The method for calculation of Aspect Maturity Index can be applied to the computation of *Class Maturity Index(CMI)*. The respective values of CMI and CIC for each version is computed using equations 8 and 9 as given below:

$$CMI = (C_c - (C_a + C_m))/C_c \quad (8)$$
$$CIC = 1 - CMI \quad (9)$$

where,

$C_c$ is the number of classes in the current release,

$C_a$ is the number of classes that were added to the current release and,

$C_m$ is the number of classes that were modified from the previous release to obtain the current release.

*2) Change in Methods(CIM)*: A method or a member function is a portion of a large application and performs a specific task. The methods have components such as, modifiers, return types, method names, parameters, an exception list and a method body. New functionalities may be added as the software evolves or the redundant functionalities may be removed. Due to this, there is a need to add, delete or modify a method. These changes are captured using equations 8 and 9 given below for measuring the *Method Maturity Index(MMI)* and *CIM* of the respective versions.

$$MMI = (M_c - (M_a + M_m))/M_c \quad (10)$$
$$CIM = 1 - MMI \quad (11)$$

where,

$M_c$ is the number of methods in the current release,

$C_a$ is the number of methods that were added to the current release and,

$M_m$ is the number of methods that were modified from the previous release to obtain the current release.

## V. EMPIRICAL EVALUATION

To measure the evolution dynamics, an open source software *AJHotDraw* has been taken as a case study and the four currently available versions have been considered. A Java based tool has been developed using *Eclipse IDE* [15] to identify and extract the elements of the base and aspect code in each version. After extraction, the tool calculates the values for the proposed set of metrics. Since the tool is used to measure the proposed metric values, it is named as *Aspect Oriented Software Evolution Measurement Tool (AOSEM Tool)*. Fig. 1 shows the overall architecture of the tool.

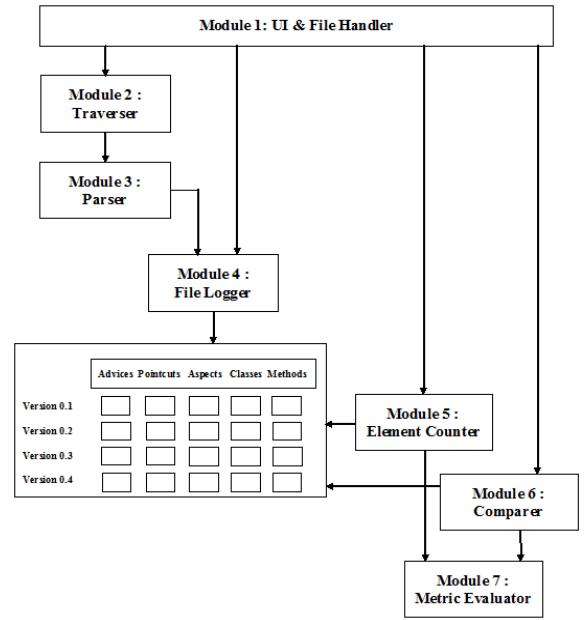

Fig. 1 Architecture of *AOSEM Tool*

*AOSEM Tool* is implemented using seven modules. *UI and File Handler* module obtains the path that contains the different versions of *AJHotDraw* as input from the user. This module is the main module and controls all the other modules of the tool. *Traverser* module is used to traverse a given folder, its directories and subdirectories in search of *.java* and *.aj* files. *Parser* module is invoked by *Traverser*. It parses the contents of each input file to separate aspects, classes, advice, methods and pointcuts. This information is stored in a separate two dimensional array. *File Writer* module is used to write the contents of the two dimensional array into separate files. A separate file is created for each version of the AO software, *AJHotDraw*.

The *Counter* module is used to count the number of entities written in the files created by the previous module. *Compare* module is invoked by the main module and reads the contents of files created for each version and compares the contents to find whether there are any changes such as addition, deletion and modification. Separate counts are maintained to reflect the number of entities added, deleted and modified when comparing two versions. Finally, the *Calculate* module is invoked by the *Compare* module to obtain the values of the proposed set of metrics. Table 1 enumerates the count of various aspect oriented entities that were measured with *AOSEM tool*.

TABLE I
COUNT OF ENTITIES ACROSS VERSIONS OF AJHOTDRAW

| Entities' Count | Version 0.1 | Version 0.2 | Version 0.3 | Version 0.4 |
|---|---|---|---|---|
| No. of Classes | 151 | 142 | 145 | 169 |
| No. of Pointcuts | 0 | 3 | 26 | 26 |
| No. of Aspects | 6 | 9 | 31 | 31 |
| No. of Advices | 1 | 6 | 42 | 42 |
| No. of Methods | 2472 | 2386 | 2432 | 2774 |

The values of the newly defined metrics are calculated using the *AOSEM Tool*. The number of classes, pointcuts, aspects and advices that were added to, deleted from or modified in the current version are also calculated and the values are tabulated in Tables 2, 3 and 4.

TABLE III
COUNT OF ENTITIES ADDED ACROSS VERSIONS OF AJHOTDRAW

| Entity | Addition to Version 0.1 | Addition to Version 0.2 | Addition to Version 0.3 |
|---|---|---|---|
| Classes | 4 | 10 | 4 |
| Pointcuts | 3 | 24 | 0 |
| Aspects | 4 | 23 | 0 |
| Advice | 5 | 37 | 0 |
| Methods | 5 | 36 | 36 |

TABLE III
COUNT OF ENTITIES DELETED ACROSS VERSIONS OF AJHOTDRAW

| Entity | Deletion to Version 0.1 | Deletion to Version 0.2 | Deletion to Version 0.3 |
|---|---|---|---|
| Classes | 13 | 7 | 4 |
| Pointcuts | 0 | 1 | 0 |
| Aspects | 1 | 1 | 0 |
| Advice | 0 | 1 | 0 |
| Methods | 0 | 0 | 0 |

TABLE IV
COUNT OF ENTITIES MODIFIED ACROSS VERSIONS OF AJHOTDRAW

| Entity | Modification to Version 0.1 | Modification to Version 0.2 | Modification to Version 0.3 |
|---|---|---|---|
| Pointcuts | 0 | 0 | 1 |
| Advice | 0 | 0 | 3 |
| Methods | 2164 | 2054 | 2054 |

VI. RESULTS AND DISCUSSIONS

It is observed that the number of entities added is more than that of the number of deleted entities. In addition, the number of entities modified is also very less compared to the number of additions. This can be visually inferred from the pie chart shown in Fig. 2. Generally, in software evolution, new requirements will be added to each version and consequently the number of entities that are added will definitely be larger than those deleted and modified. A fewer number of entities may be deleted to optimize the code or to reduce the redundancy.

TABLE V
MATURITY METRIC VALUES FOR VERSIONS OF AJHOTDRAW

| Metric | Version 0.1 | Version 0.2 | Version 0.3 | Version 0.4 |
|---|---|---|---|---|
| AMI | 1 | 0.5556 | 0.2581 | 1 |
| PMI | 1 | 0 | 0.0769 | 0.9615 |
| ADMI | 1 | 0.1667 | 0.1190 | 0.9286 |
| CMI | 1 | 0.9718 | 0.9310 | 0.9763 |
| MMI | 1 | 0.0909 | 0.1406 | 0.2466 |

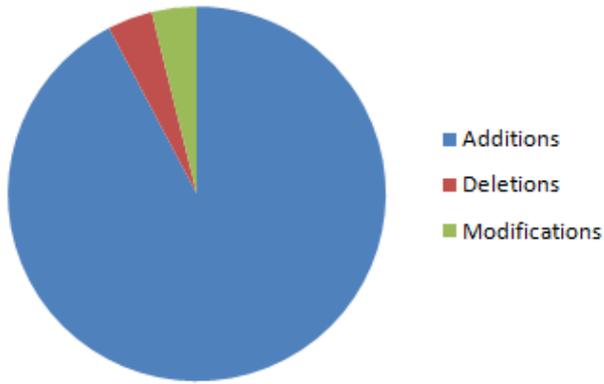

Fig. 2 Pie Chart depicting the spread for total of Change types across all versions of *AJHotDraw*

TABLE VI
CHANGE METRIC VALUES FOR THE VERSIONS OF AJHOTDRAW

| Metric | Version 0.1 | Version 0.2 | Version 0.3 | Version 0.4 |
|---|---|---|---|---|
| *CIA* | 0 | 0.4444 | 0.7419 | 0 |
| *CIP* | 0 | 0 | 0.9231 | 0.0385 |
| *CIAD* | 0 | 0.8333 | 0.8810 | 0.0714 |
| *CIC* | 0 | 0.0282 | 0.0690 | 0.0237 |
| *CIM* | 0 | 0.9091 | 0.8594 | 0.7534 |

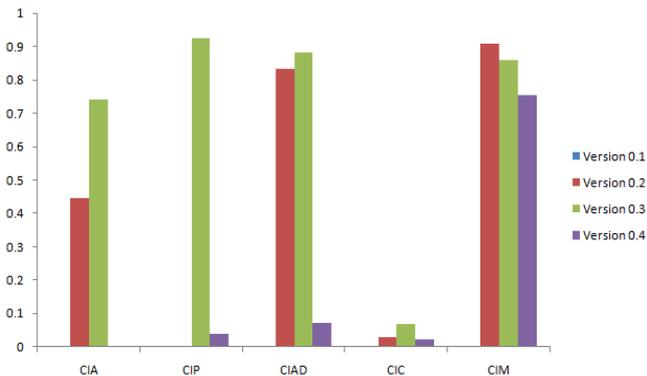

Fig. 3 Comparison on the metric values for different versions of *AJHotDraw*

The graph in Fig. 3 visually explains the comparison among values of the newly defined metrics across versions. From the graph, it can be observed that the metric values for version 0.2 are higher than those of version 0.3. This clearly shows the dynamics of evolution over the four versions of *AJHotDraw*.

The following inferences have been drawn from the case study that considered four different versions of *AJHotDraw*:

In version 0.2, there is a major change in the values of *CIP* (Change in Pointcuts) and *CIAD* (Change in advices). This is because the aspects are introduced only in version 0.2. The value of *CIAD* has steadily increased in the subsequent version (version 0.3). In version 0.1, the persistence crosscutting concern was not part of the tool and was developed as a test suite using classes. It is only in version 0.2, that the persistence crosscutting concern is implemented through aspectual elements. In version 0.3, the other cross-cutting concerns like *observer pattern* instance and *undo* were also abstracted as aspects. Hence, the number of aspect specific elements (like aspects, pointcuts, and advices) is less in version 0.1, when compared to the subsequent versions. This shows that the aspect specific elements are increased during the evolution of software. The initial versions used aspects to abstract the non-functional requirements (persistence), while the subsequent versions modeled the functional requirements (observer pattern instance, undo, cut and paste) also as aspects. This is evident from the observation that there is a good change in the values of *CIP* and *CIAD* for version 0.3.

The change in the value of *CIA* (Change in Aspects) increases over versions. The number of aspects added to version 0.3 is more than that of version 0.2 and when it comes to deletion, the value is minimal. In version 0.3, no aspect has been added or deleted. This shows that a good number of cross-cutting concerns are modeled as aspects in version 0.4. It can also be said that almost all the cross-cutting functionalities of the matured version of *AJHotDraw* have also been abstracted as aspects and evolution of aspects in *AJHotDraw* has reached a maximum degree of maturity with respect to the corresponding set of requirements.

The value of *CIM* (Change in Methods) is high in version 0.2 and there is a decline in version 0.3. This might be because in version 0.2, only the persistence functionality is refactored as aspect and the rest being refactored in version 0.3. In version 0.4, very minimal modifications are done to the existing modules resulting in even lower value of *CIM*. The functionalities which are cross-cutting are slowly being moved into aspects leading to a decline in the value of *CIM*. This is also evident from the proportional increase in the value of *CIAD* across versions.

There is not much change in the value of *CIC* (Change in Classes) in all the three versions. The value of *CIM* (Change in Methods) also reduces over versions. This shows that the cross-cutting concerns previously modeled as classes and methods are now embedded within aspects. This is also evident from the increase in the value of *CIA* and *CIAD* across versions.

## VII. CONCLUSIONS AND FUTURE DIRECTIONS

A group of metrics have been defined which were used to measure the different elements of the software, *AJHotDraw* developed using AO methodology. The metrics were used to measure the software evolution, an important quality attribute for the sustenance of any successful software. These metrics also capture the evolution dynamics of the AO software. It was found that there were more additions of classes and aspects compared to deletions and modifications. Further, in the latest version of *AJHotDraw*, the functional concerns were also modeled as aspects. As an extension of this work, more case studies of AO software can be analyzed using the set of metrics defined in this paper. The inferences can be generalized to find the impact of evolution in AO based applications.

Further, additional metrics can be defined to capture introductions in an aspect and changes in the granular level, say fields and precedence. The concern diffusion metrics cited by Sant`Anna [13] can also be used to capture the evolution of AO software over versions. All these metrics can be collectively used to study the overall impact on software quality.